\documentclass[11pt]{article}
\usepackage{amsfonts}
\usepackage{amsmath}
\usepackage{amsthm}
\usepackage{amssymb}
\usepackage{amscd}

\def\be{\begin{equation}}
\def\ee{\end{equation}}

\def\bea{\begin{eqnarray}}
\def\eea{\end{eqnarray}}

\def\a{\alpha}
\def\b{\beta}
\def\c{\gamma}
\def\d{\delta}
\def\C{\Gamma}
\def\D{\Delta}
\def\e{\epsilon}
\def\p{\varphi}
\def\P{\Phi}

\def\w{\omega}

\def\rs{r_{*}}
\def\tnabla{\tilde{\nabla}}

\begin{document}

\begin{flushright}
DAMTP-2002-75
\end{flushright}

\vspace{3cm}

\begin{center}
{\LARGE {\bf Gravitational instability in higher dimensions}}
\vspace{1cm}

Gary Gibbons and Sean A. Hartnoll \\
\vspace{0.3cm}

{\it DAMTP, Centre for Mathematical Sciences, Cambridge University,\\
Wilberforce Road, Cambridge CB3 0WA, UK.}
\vspace{0.5cm}

\noindent G.W.Gibbons@damtp.cam.ac.uk \hspace{1cm} S.A.Hartnoll@damtp.cam.ac.uk
\end{center}
\vspace{1cm}

\begin{abstract}
We explore a classical instability of spacetimes
of dimension $D>4$. Firstly, we consider static solutions:
generalised black holes and brane world metrics.
The dangerous mode is a tensor mode
on an Einstein base manifold of dimension $D-2$.
A criterion for instability is found for the generalised Schwarzschild,
AdS-Schwarzschild and topological black hole spacetimes in terms of
the Lichnerowicz spectrum on the base manifold.
Secondly, we consider perturbations in time-dependent solutions:
Generalised dS and AdS. Thirdly we show that, subject to the usual
limitations of a linear analysis, any Ricci flat spacetime may be
stabilised by embedding into a higher dimensional spacetime
with cosmological constant. We apply our results to pure AdS black strings.
Finally, we study the stability of higher dimensional ``bubbles of
nothing''.

\end{abstract}

\pagebreak
\pagenumbering{arabic}

{\small
\tableofcontents
}

\section{Introduction}

Over the last few years solutions of the Einstein equations in
higher dimensions have come to play an important role as
background metrics in various physical applications.
These range from theories of TeV gravity, where higher dimensional
black holes are predicted to be produced in the next
generation of colliders \cite{dl,gt}, to the gravity-gauge theory
correspondence \cite{mal,agmoo}.

Clearly the stability of such spacetimes is an important issue.
One feature of higher dimensional spacetimes is that they
often satisfy boundary conditions which
differ from those encountered in four spacetime dimensions.
This is because the two-sphere and two dimensional
hyperbolic space are, up to discrete quotients, the unique Einstein
manifolds in two dimensions with positive and negative curvature
respectively. In higher dimensions there are more possibilities
\cite{gis, gis2}. These include metrics such as the Bohm metrics \cite{bohm}
that exist on manifolds that are topologically $S^d$.

In particular, we will consider the case in which the higher
dimensional spacetime includes a $d$-dimensional Einstein manifold,
$\{B,\tilde{g}\}$, which we call the base manifold, in a common
way. In these cases we shall show that part and sometimes all of the
stability problem may be reduced to the solution of an ordinary
differential equation of Schr\"odinger form.
The modes we concentrate on are transverse tracefree tensor
harmonics on the base manifold, $\{B,\tilde{g}\}$.
The differential equation determining stability of the spacetime
then depends on the spectrum of the Lichnerowicz operator on
transverse traceless symmetric tensor fields of the manifold $B$.
These modes do not exist in the stability analysis of, for example, the
Schwarzschild black hole in four dimensions \cite{rw,vish,zer,wald}
because there are no suitable tensor harmonics on $S^2$ \cite{hig}. Thus,
the instabilities we discuss are inherently higher dimensional.

Typically, the metric $\tilde{g}$ on $B$ will be such that
\be\label{eq:econst}
\tilde{R}_{\a\b}=\e (d-1) \tilde{g}_{\a\b},
\end{equation}
with $\e = \pm 1$ or $\e = 0$.
This is the normalisation of $S^d$, for example.
Tildes denote tensors on $B$.

Some examples of such spacetimes are now in order. 

\vspace{0.5cm}

\noindent {\bf (a) Static solutions}

The spacetime is $D=d+2$ dimensional and of the form
\be
ds^2 = - f(r) dt^2 + \frac{dr^2}{f(r)} + r^2 d\tilde{s}^2_d 
\end{equation}
with
\be
f(r) = \e-\left(\frac{\a}{r}\right)^{d-1} - cr^2 ,
\end{equation}
and $d\tilde{s}^2_d$ is the metric on $B$.
The cosmological constant in $d+2$ dimensions is
\be
R_{ab} = c (d+1) g_{ab} .
\end{equation}

Consider first vanishing cosmological constant, $c=0$ and with $\e=1$ in
(\ref{eq:econst}). When
$B=S^d$, these are the Schwarzschld-Tangherlini black holes
\cite{tan} which are spatially Asymptotically Euclidean (AE). If $B\neq S^d$ one
obtains generalised higher dimensional black holes \cite{gis} which
are spatially Asymptotically Conical (AC). These are of course not possible in
four dimensions because $S^2$ is the only positive curvature Einstein
manifold in two dimensions.
If $B = S^d/\Gamma$ where $\Gamma \subset SO(d+1)$ is discrete,
then the spatial metric will be Asymptotically Locally Euclidean (ALE). The
$(d+1)$-dimensional Riemanian manifold with metric
\be
d\rho ^2 + \rho ^2 d\tilde{s}^2_d, \label{eq:cone}
\end{equation}
is called the cone $C(B)$ with base $B$.
It is Ricci flat precisely if $\tilde{g}$ is Einstein with the
Einstein constant of (\ref{eq:econst}). If $B \ne S^d$ there will
typically be a singularity at
the vertex of the cone, but in our case this will often be hidden inside
an event horizon. 

In forming time dependent solutions below we will replace $\rho ^2$
by $\sin^2 \rho$ or
$\sinh ^2 \rho$ in (\ref{eq:cone}) to obtain $(d+1)$ dimensional Einstein
metrics of positive or negative scalar curvature respectively. This
will have two or one singular vertices respectively. In the
latter case the Riemannian metric will be Asymptotically
Hyperbolic (AH). Another way to get an Asymptotically Hyperbolic Einstein
metric with negative scalar curvature is to take $\epsilon=0$ and
replace $\rho ^2$ by $e^{2 \rho }$.

If $c$ is positive and $\epsilon=1$, one gets a
generalized Schwarzschild-Tangherlini-de Sitter spacetime. The
static region between a cosmological event horizon and a black
hole event horizon is non-singular. If $c$ is negative and
$\epsilon=1$, one has generalised
Schwarzschild-Tangherlini-Anti-de Sitter, without a cosmological horizon.

Another interesting possibility with no cosmological horizon
is to take $c$ negative
and $\epsilon=-1$ \cite{van,bir,mann}. Now, if $\a=0$, the resulting metric will
be singularity free. If $B$ is a hyperbolic manifold $B=
H^d/\Gamma$, with $\Gamma \subset SO(d,1)$ a suitable discrete
group, then we have an identification of Anti-de Sitter spacetime
sometimes thought of as a topological black hole. However if $B$ is
not a hyperbolic manifold then one gets a singularity free
topological black hole which is not locally isometric to
Anti-de Sitter spacetime.

Also of this form are solutions with negative cosmological constant of
a sort which arise in brane world scenarios \cite{rs1, rs2}. The metric is most
familiar in the form
\be
ds^2 = \frac{1}{z^2} \Bigl ( dz^2 - dt ^2 + d\tilde{s}^2_d  
\Bigr ) ,
\end{equation}
with $\e=0$. If $B$ is flat we obtain $d+2$
dimensional Anti-de Sitter spacetime.

\vspace{0.5cm}

\noindent{\bf (b) Time-dependent solutions}

By suitably re-interpreting our
formulae we can also discuss the stability of
some time-dependent solutions. For example a generalized $D=d+1$
dimensional de Sitter spacetime is given by
\be
ds^2 = -dt ^2  + \frac{\cosh^2 Lt}{L^2} d\tilde{s}^2_d. \label{eq:Sitter}
\end{equation}
with $\epsilon=1$.  This is singularity free.
Changing $\cosh Lt$ to $\sin Lt$ in (\ref{eq:Sitter}) and letting $\e=-1$
will give a generalized Anti-de Sitter
spacetime which will have Big Bang and Big Crunch singularities
at $t=0$ and $t= \pi/L$ respectively unless $B$ is the hyperbolic metric
on $H^d$.

\vspace{0.5cm}

\noindent {\bf (c) Ricci flat Lorentzian base and double analytic continuation}

A simple generalization of this time dependent situation
arises if we take $B$ to be a $d$ dimensional Lorentzian
Ricci flat manifold whose stability properties are known.

For example we could consider the $d+1$ dimensional Einstein
manifold with negative scalar curvature whose metric is
\be ds^2=
\frac{1}{z^2} \Bigl ( dz^2 + d\tilde{s}^2_d \Bigr ),
\end{equation}
with $\epsilon =0$, because the base is Ricci flat.
If $B$ is the flat Minkowski metric then we have the metric of
$d+1$-dimensional Anti-de Sitter spacetime. If $B$ is a black hole
metric, then we have black strings in Anti-de Sitter
spacetime.

Another situation in which a Lorentzian base arises is in double analytic
continuation of black hole metrics. The resulting solutions describe
expanding ``bubbles of nothing''\cite{wit}. Double analytic continuation of
generalised Schwarzschild gives the metric
\be
ds^2 = \left[ 1 - \left( \frac{\a}{r} \right)^{d-1} \right] d\psi^2 +
\frac{dr^2}{1 - \left( \frac{\a}{r} \right)^{d-1}} + r^2 d\tilde{s}^2_{d},
\end{equation}
where $\psi$ is periodic and $d\tilde{s}^2_{d}$ is a Lorentzian metric obtained
via analytic continuation of a Euclidean Einstein metric with $\e=1$.
If the Euclidean base is $S^d$, then the corresponding Lorentzian base
is just de Sitter space, $dS_d$.

In \S 2 we relate the Lichnerowicz operator on certain modes in the spacetime
with the Lichnerowicz operator on the base manifold. This will give us equations
for the perturbative modes. In \S 3 we study the stability of generalised static metrics by setting up
a Sturm-Liouville problem. In \S 4 we look at perturbations in time dependent scenarios.
In \S 5 we recall the Lichnerowicz spectra on some manifolds that give explicit examples for
the results of section 3. Finally, \S 6 contains the conclusions.

\section{Lichnerowicz operator on a class of spacetimes}

\subsection{A Lichnerowicz mode}

Consider a $D$ dimensional spacetime with metric
\be\label{eq:metric}
ds^2_D = - f(r) dt^2 + g(r) dr^2 + r^2 d\tilde{s}^2_d ,
\end{equation}
where $d\tilde{s}^2_d$ is a Riemannian metric
on a $d=D-2$ dimensional manifold $B$.
The spacetime is taken to be Einstein.

The Lichnerowicz operator acting on a symmetric second rank tensor $h$
is
\be\label{eq:lich}
(\D_L h)_{ab} = 2 R^c{}_{abd} h^d{}_c + R_{ca} h^c{}_b + R_{cb}
h^c{}_a - \nabla^c \nabla_c h_{ab} .
\end{equation}
For transverse tracefree perturbations this gives the first order change
in the Ricci tensor under a small perturbation to the metric
\bea\label{eq:gauge}
& & g_{ab} \to g_{ab} + h_{ab}, \quad\mbox{such that}\quad h^a{}_a
= \nabla^a h_{ab} = 0 \nonumber \\
& & R_{ab} \to R_{ab} + \frac{1}{2} (\D_L h)_{ab} .
\eea
The Lichnerowicz operator is compatible with the transverse,
tracefree condition \cite{gp}.

We wish to study the stability of metrics of the form (\ref{eq:metric})
under certain metric perturbations. It will be useful to have an
expression for the Lichnerowicz operator on the spacetime in terms of
the Lichnerowicz operator on the base manifold
$B$. We shall impose the conditions
\be\label{eq:extra}
h_{0a} = h_{1a} = 0 ,
\end{equation}
where $0,1$ are the $t,r$ coordinates. These conditions are of course
not a gauge choice and mean
that we are restricting the modes we are looking at.
More precisely, we are restricting attention to tensor modes on the base
manifold and we are not considering scalar and vector modes.
However, following \cite{dfghm} we argue in appendix A that, at least
for the manifolds in which the base is compact and Riemannian with $\e=1$,
the stability of the spacetime under scalar and vector perturbations
is insensitive to the base manifold. Therefore, for these modes
one may consider the base to be the sphere, $S^d$. But this leaves us with
just the standard Schwarzschild-Tangherlini(-AdS) spacetimes. These
standard higher dimensional black holes are expected, although to our
knowledge not proven, to be stable against vector and scalar perturbations.
Therefore, we expect that it is only the tensor modes which probe the base manifold
sufficiently to produce instabilities. Nonetheless, it would be nice to see this
from an explicit perturbation analysis.
The conditions (\ref{eq:extra}) and the form of the metric (\ref{eq:metric})
imply that the transeverse tracefree property of $h_{ab}$
(\ref{eq:gauge}) is inherited by $h_{\a\b}$.
Here and throughout the indices
$a,b,\ldots$ run from $0\ldots D$ and the indices $\a,\b,\ldots$ will run from
$2\ldots D$ and are the coordinates on $B$. 

A calculation then gives
\bea\label{eq:dl}
(\D_L h)_{\a\b} & = & \frac{1}{r^2}(\widetilde{\D}_L h)_{\a\b}
+ \frac{1}{f} \frac{d^2}{dt^2} h_{\a\b} - \frac{1}{g} \frac{d^2}{dr^2}
h_{\a\b} \nonumber \\
& + &  \left[\frac{-f^{\prime}}{2fg} + \frac{g^{\prime}}{2g^2} +
\frac{4-d}{gr} \right] \frac{d}{dr} h_{\a\b} - \frac{4}{gr^2} h_{\a\b} ,
\eea
where $(\widetilde{\D}_L h)_{\a\b}$ is the Lichnerowicz
operator on $B$. All the other components of $(\D_L
h)_{ab}$ are zero because of the transverse tracefree property.
This expression is the backbone of all the calculations
in this paper.

It should be noted that (\ref{eq:extra}) are strong conditions in low
dimensions. For the four dimensional Schwarzschild solution, for
example, there are no perturbations of this form because $S^2$ does
not admit any tensor harmonics \cite{hig}. 
{\it We are looking at a
potentially unstable mode that is specific to higher dimensional spacetimes.}

Our strategy in applying this to static spacetimes in the next section
will be to calculate first a criterion for instability
in terms of the minimum Lichnerowicz eigenvalue, $\lambda_{min}$,
on the base manifold $B$. In a later section we will then find this minimum
for several relevant manifolds.

\subsection{Gauge Freedom}

Diffeomorphism invariance of the Einstein equations implies
gauge invariance of the linear theory under
\be
h_{ab} \to h_{ab} + \nabla_a \xi_b + \nabla_b \xi_a .
\end{equation}
The invariance can be used to set
\be\label{eq:trans}
\nabla^{a} \bar{h}_{ab} \equiv \nabla^{a} (h_{ab} - \frac{1}{2} g_{ab} h^c{}_c) = 0 .
\end{equation}
This is the transverse gauge condition which may always be imposed.
The residual gauge freedom is given by vectors $\xi$ satisfying
\be\label{eq:resid}
\square \xi_a + R_a{}^b \xi_b = 0 .
\end{equation}
Recall that the trace transforms as $h^a{}_a \to h^a{}_a + 2 \nabla^a \xi_a$, we
would like to find a $\xi$ satisfying (\ref{eq:resid}) and such that $h^a{}_a \to 0$.

We show in appendix B that if the background spacetime is vacuum, possibly with a
cosmological constant, then one may impose the transverse tracefree condition
for perturbations as a gauge choice. This is slightly more subtle than
the standard argument in which the cosmological constant is zero.

Therefore the transverse tracefree choice made in the previous subsection is merely a gauge choice
if the background spacetime is vaccum, with or without a cosmological constant. However, 
this will not fix all the gauge freedom and we need to check that any solutions we find are not pure gauge.
A pure gauge solution would be of the form
\be
h_{ab} = \nabla_a \xi_b + \nabla_b \xi_a .
\end{equation}
In appendix C we show that none of the modes considered in this paper
are pure gauge.

\section{Application to static metrics}

\subsection{Sturm-Liouville problem}

In this section we consider static metrics which solve the
vacuum Einstein equations, possibly with a cosmological constant
\be
R_{ab} = c (d+1) g_{ab} .
\end{equation}
This requires $g=1/f$
and the metric on $B$ will be Einstein with
\be
\tilde{R}_{\a\b}=\e (d-1) \tilde{g}_{\a\b},
\end{equation}
with $\e = \pm 1$ or $\e = 0$.
Tildes denote tensors on $B$.
The cases $\e = \pm 1$
correspond to having the same scalar curvature as $S^d$
or $H^d$.
The function $f$ must be of the form
\be\label{eq:fform}
f(r) = \e-\left(\frac{\a}{r}\right)^{d-1} - cr^2.
\end{equation}

We look for unstable modes of the form
\be\label{eq:mode}
h_{\a\b}(x) = \widetilde{h}_{\a\b}(\tilde{x}) r^2 \p(r) e^{\w t} ,
\end{equation}
where $\tilde{x}$ are coordinates on $B$ and
\be
(\widetilde{\D}_L \tilde{h})_{\a\b} = \lambda \widetilde{h}_{\a\b} .
\end{equation}
As discussed in the previous section and in appendix A, we expect these
tensor modes to be the dangerous modes. This is similar to the situation
encountered in recent studies of stability
of $AdS_p \times M_q$ metrics \cite{dfghm,gm,siot}.

The pertubation must satisfy $\d R_{\a\b}=c(d+1) h_{\a\b}$ and this gives an
equation for $\varphi$ that may be cast in Sturm-Liouville form
\be\label{eq:sl}
- \frac{d}{dr} \left( fr^d\frac{d\p}{dr}\right) +
\left(\frac{\lambda}{r^2} - \frac{2 f^{\prime}}{r} - \frac{(2d-2)f}{r^2} - 2c(d+1)
\right) r^d \p = -\w^2 \frac{r^d}{f}\p .
\end{equation}
It is convenient to rewrite this as a Schr\"odinger equation by
changing variables to Regge-Wheeler type coordinates and rescaling
\be\label{eq:change}
d\rs = \frac{dr}{f} ,\quad\quad \P = r^{\frac{d}{2}} \p .
\end{equation}
Equation (\ref{eq:sl}) now becomes
\be\label{eq:schro}
- \frac{d^2\P}{d\rs^2} + V(r(\rs)) \P = - \w^2 \P \equiv E \P,
\end{equation}
where the potential is
\be\label{eq:V}
V(r) = \frac{\lambda f}{r^2} + \frac{d-4}{2}\frac{f^{\prime}f}{r} + \frac{d^2-10d+8}{4}\frac{f^2}{r^2}
- 2c(d+1) f.
\end{equation}
Thus the stability problem reduces to the existence of bound states with $E < 0$
of the Schr\"odinger equation with potential $V(r)$.
If such a bound state of the Schr\"odinger equation
exists then the spacetime (\ref{eq:metric}) is unstable to modes
of the form (\ref{eq:mode}). That is to say, there will be an instability if the
ground state eigenvalue, $E_0$, of (\ref{eq:schro}) is negative.

The normalisation of wavefunctions must take into account the weight function of (\ref{eq:sl}),
but the usual normalisation is recovered for $\P$
\be\label{eq:normal}
1 = \int \p^2 \frac{r^d}{f} dr = \int \P^2 \frac{dr}{f} = \int \P^2 d\rs .
\end{equation}
This condition of normalisability that is necessary to set up the Sturm-Liouville problem
is just the condition of finite energy of the gravitational
perturbation (\ref{eq:mode}). The background spacetimes (\ref{eq:metric}) have
a timelike Killing vector $\xi^0 = 1$, up to consideration of horizons.
This allows the total energy of the perturbation on a spacelike hypersurface with normal
$n^0 = 1/f^{1/2}$ to be well-defined, independently of
the asymptotics of the background spacetime \cite{ad}
\be\label{eq:ener}
E \propto \int t^{\mu \nu} n_{\mu}\xi_{\nu}\sqrt{g^{(d+1)}} d^{d+1}x =
\int \frac{r^d}{f} t_{00} \sqrt{\tilde{g}^{(d)}} d^d\tilde{x} dr ,
\end{equation}
where $t_{00} = G_{00}^{(2)} = R_{00}^{(2)} - \frac{1}{2} g_{00} R^{(2)}$, the second
order change in the Einstein tensor under the perturbation. Now note that the kinetic part
of (\ref{eq:ener}) contains terms like
\bea
\int \frac{r^d}{f} h^{\a\b}\,\partial_0\,\partial_0\, h_{\a\b} \sqrt{\tilde{g}^{(d)}}  d^d\tilde{x} dr
& \propto & e^{2\w t} \int \tilde{h}^{\a\b} \tilde{h}_{\a\b} \sqrt{\tilde{g}^{(d)}} d^d\tilde{x} 
\int \p^2 \frac{r^d}{f} dr \\ \nonumber
& \propto & \int \p^2 \frac{r^d}{f} dr .
\eea
Thus requiring finite energy will recover the normalisation (\ref{eq:normal}).

Besides normalisation, we must also consider boundedness properties. The linear approximation
to the equations of motion requires that $h^a{}_b \ll 1$. For black hole
spacetimes with event horizons, one should re-express solutions in Kruskal coordinates \cite{vish,gl3}
near the horizon and check boundedness there. This is because Kruskal coordinates are well-behaved
at the horizon. Fortunately, the mode we are considering has
no $t$ or $r$ components and therefore is essentially unchanged in Kruskal coordinates. Thus it is
sufficient to check boundedness in $r$ in the original coordinates of (\ref{eq:metric}). However,
in \S 3.2.2 we show explicitly boundedness in Kruskal coordinates for completeness.

From (\ref{eq:mode}), we see that boundedness requires
$\p(r) = \P(r) r^{-\frac{d}{2}} \ll 1$. As $r\to\infty$ the function
$\P(r)$ must go as $r^{\frac{d}{2}}$ or lower power of $r$. This is a weaker constraint than is imposed by
finite energy (\ref{eq:normal}), with or without cosmological constant. As $r\to 0$ we must
have $\P(r)$ going as $r^{\frac{d}{2}}$ or higher power of $r$. This will almost always be a stronger constraint than
that required by finite energy (\ref{eq:normal}).
However, in many of the applications in this section
there will be an event horizon at some finite $r_0$, where $f(r_0)=0$. The condition of boundedness will then
simply be that $\P(r)$ is bounded at $r_0$. The finite energy condition will be that $\P(r)$ goes to zero
on the horizon, because the zero of $f(r)$ is simple. In the cases below, we need to impose the stronger
condition for each limit. However, in almost all cases we encounter,
the soutions either satisfy both or neither of the criteria.

\subsection{Vanishing cosmological constant}

\subsubsection{Asymptotic criterion for stability}

Set the cosmological constant $c=0$. Asymptotically $f\to 1$ and
$r=\rs$. Alternatively, this is the massless case.
We will derive first a criterion for instability by solving the
asymptotic Schr\"odinger equation (\ref{eq:schro}) with $f=1$, and hence
\be\label{eq:vinf}
V_{\infty}(r)=\frac{d^2-10d+8+4\lambda}{4r^2} ,
\end{equation}
and requiring suitable behaviour in the interior. Call the asymptotic solution $\P_{\infty}$.
The range here is $0\leq r < \infty$. The argument of this subsection is in
fact independent of the interior form of $f$.

The asymptotic solution which decays at infinity is
\be\label{eq:bessel}
\P_{\infty}(r) = \Re\left[ r^{\frac{1}{2}} K_{\nu}(\w r)\right],
\quad\quad\nu = \frac{1}{2}\sqrt{(5-d)^2-4(4-\lambda)} ,
\end{equation}
where $K_{\nu}(\w r) $ is the modified Bessel function that decays at
infinity \cite{spec}. The behaviour of (\ref{eq:bessel}) for small $r$ and real, positive $\nu$ is
$\P_{\infty}(r) \sim r^{-\nu+1/2}$. Three cases should be distinguished. If
$\nu \geq 1$ the solution is divergent and not normalisable according to (\ref{eq:normal}).
If $1 > \nu > 1/2$, the solution is divergent but normalisable. If
$1/2 > \nu \geq 0$, the solution goes to zero for small $r$. Another possibility is that
the index $\nu=i\nu_i$ is
pure imaginary, in which case the Bessel function oscillates in the interior as $\sin(\nu_i\ln r)$
and the wavefunction $\P$ is then normalisable.
We see that non-divergent normalisable solutions occur precisely when the potential (\ref{eq:vinf})
is negative, that is
\be
\frac{d^2-10d+8+4\lambda}{4} = \nu^2 -
\frac{1}{4} \leq 0 .
\end{equation}
We have finite energy solutions for
a continuous range of $\w>0$. The continuous spectrum of arbitrarily low energy
is a direct consequence of the asymptotic potential (\ref{eq:vinf}) being unbounded below.
This will not be the case for the full potential and the spectrum will become discrete.

In all of the cases of the previous paragraph,
$\p = \P r^{-\frac{d}{2}}$ is not bound at the origin and so none of these
solutions give instabilities of the massles, $f=1$, metric. However, 
following \cite{lo} we note that in the oscillatory solutions with imaginary index,
the derivative takes all values and so one might expect to be able to match
the asymptotic solution to an interior solution for which $f\neq 1$, at least
for certain discrete values of $\w$. Thus
if a $\lambda$ exists such that $\nu$ is imaginary then the Schr\"odinger
equation should have a bound state and the metric is unstable. That is
\be\label{eq:crit}
\lambda_{min} < \lambda_c \equiv 4 - \frac{(5-d)^2}{4} \quad\Leftrightarrow\quad \mbox{instability} .
\end{equation}
This is the criterion for instability of a massive black hole. We have also shown that the massless
case is always stable. Concrete examples of Lichnerowicz spectra
giving stable and unstable spacetimes are given in \S 5.

The vacuum solution for $f$ is of course (\ref{eq:fform}) which is now just the
Schwarzschild-Tangherlini \cite{tan} black hole and the Asymptotically Conical (AC) variants
considered in \cite{gis,gis2}. The radial function is
\be
f(r) = 1-\left(\frac{\a}{r}\right)^{d-1} .
\end{equation}
The higher dimensional Regge-Wheeler tortoise coordinate (\ref{eq:change}) may be given
explicitly in this case as \cite{gl3}
\be\label{eq:rw}
\rs = r + \sum_{n=1}^{d-1} \frac{e^{2\pi i n/(d-1)}}{d-1} \a \ln \left( r - e^{2\pi i n/(d-1)} \a\right) .
\end{equation}
There is an event horizon at $r=\a$, and so the range of $r$ is $\a\leq r$.
The potential becomes
\be\label{eq:potsc}
V(r) = V_{\infty}(r) + \frac{1}{\a^2} \left(\frac{\a}{r}\right)^{d+1} \left[ \frac{10d-8-4\lambda}{4}
-\left(\frac{\a}{r}\right)^{d-1} \frac{d^2}{4} \right] .
\end{equation}
It follows that $V(\a) = 0$, as was clear from the initial definition (\ref{eq:V}), and that $V(r) \to 0$ as $r\to\infty$.

The potential (\ref{eq:potsc}) can be seen to be always positive for $\a\leq r <\infty$
if $d^2-10d+8+4\lambda \geq 0$. This was
the condition for the asymptotic potential to be positive also and signalled the nonexistence of
finite energy solutions $\P$ in this case. Thus, as expected, there are also no solutions to the full
equations in this case.

To establish the criterion for instability (\ref{eq:crit}) we still need to check that there are solutions
when the asymptotic solution oscillates in the interior and that there are none for the range
$1/2 \geq \nu \geq 0$, where an asymptotic solutions exists but does not oscillate. These statements
will be supported numerically in the next section. The conclusion will be that {\it a generalised black hole
is unstable if and only if the base manifold has a Lichnerowicz spectrum satisfying (\ref{eq:crit})}.

\subsubsection{Numerical support for the asymptotic criterion}

The Schr\"odinger equation (\ref{eq:schro}) with potential (\ref{eq:potsc}) has a regular singular point
at the event horizon $r=\a$. Thus we may perform a Taylor expansion of the solution of the equation
about this point. The leading order terms are found to be $\P \sim
(r-\a)^{\pm \a\w/(d-1)}$, so long as the exponent is non-integer. Typically there
is thus one divergent and one convergent solution at the horizon. Further, the convergent
solution vanishes on the horizon and therefore satisfies both the finite energy and boundedness requirements.
We would like to see whether the
solution that is well behaved at the horizon is also well behaved at infinity, giving a bound state.

Before solving the equation we can check explicitly, following \cite{gm3}, 
that the well behaved solution remains well behaved at the
horizon in Kruskal coordinates. Including the time dependence and using the limit of (\ref{eq:rw}) as
$r\to\a$, we see that the mode (\ref{eq:mode}) behaves near the horizon as
\be
h_{\a\b} \sim (r-\a)^{\a\w/(d-1)} e^{\w t} \tilde{h}_{\a\b}
= e^{\w\left(t+\a(r-\a)/(d-1)\right)}\tilde{h}_{\a\b} \sim e^{\w(t+\rs)}\tilde{h}_{\a\b} .
\end{equation}
The Kruskal coordinates $R,T$ are given by
\be
R \pm T \sim e^{f^{\prime}(\a)(\rs\pm t)/2} = e^{(d-1)(\rs\pm t)/2\a}.
\end{equation}
Therefore the mode goes as
\be
h_{\a\b} \sim (R+T)^{2\w/f^{\prime}(\a)}\tilde{h}_{\a\b} ,
\end{equation}
which is well behaved on the future horizon $R-T=0$. It is not difficult to see that this
expression will remain true for other functions $f(r)$, such as that for the AdS black holes studied
below. In this case $f$ should of course be evaluated at the horizon, which would no longer be $\a$.

To investigate the equation numerically we first find a series expansion about the horizon of the solution
that is regular at the horizon. We use this to
set the initial conditions away from horizon itself. By taking sufficient terms in the series, this may be done
to high accuracy. We then choose a (positive) value for $\w$ and the equation can be numerically integrated.
The solution will always diverge for large $r$ because it is extremely unlikely that the $\w$ we have
specified corresponds precisely to a bound state. However, by varying $\w$ we may see that the solution
diverges to positive infinity for some values of $\w$ and to negative infinity for others. Because the solutions
of the differential equation depend continuously on the parameters of the equation, there must be
a bound state for some intermediate value of $\w$. The solutions can be double checked by integrating
in from infinity towards the horizon, although this is less accurate
when $\w$ is small, as for the interesting cases.
Tables 1 and 2 show values of $\w$ between which
the lowest lying negative energy bound state, $\w_{max}^2 = - E_0$, is found for various small values of $d$ and
$\lambda_c - \lambda$, where $\lambda_c$ is the critical value of $\lambda$ of (\ref{eq:crit}).
Without loss of generality we take $\a=1$. The dependence of the eigenvalues on $\a$ is
determined on dimensional grounds to be $\w \propto 1/\a$. Another way of seing this is that
the Schr\"odinger equation is invariant under $\a\to k\a$, $r\to kr$, $\w\to\w /k$.

\begin{table}[h]
{\bf Table 1: } Lowest bound states for $\a=1,d=4$
  ($D=6$) generalised Schwarzschild. \\
  \begin{tabular}{|c|c|c|} \hline
$\lambda_c - \lambda$ & Lower bound for $\w_{max}$ & Upper bound for $\w_{max}$ \\ \hline \hline
$0.5$ & $3.63\times 10^{-2} $ & $3.654\times 10^{-2}$ \\ \hline
$0.2$ & $2.64\times 10^{-3} $ & $2.67\times 10^{-3} $ \\ \hline
$0.1$ & $1.41\times 10^{-4}$ & $1.416\times 10^{-4} $ \\ \hline
$0.08$ & $4.341\times 10^{-5}$ & $4.35\times 10^{-5} $ \\ \hline
$0.05$ & $2.22\times 10^{-6}$ & $2.28\times 10^{-6} $ \\ \hline
$< 0$ & \multicolumn{2}{|c|}{No solutions found.} \\ \hline
  \end{tabular}
\end{table}

\begin{table}[h]
{\bf Table 2: } Lowest bound states for $\a=1,d=8$ ($D=10$) generalised Schwarzschild.\\
  \begin{tabular}{|c|c|c|} \hline
$\lambda_c - \lambda$ & Lower bound for $\w_{max}$ & Upper bound for $\w_{max}$ \\ \hline \hline
$0.5$ & $2.28\times 10^{-2} $ & $2.29\times 10^{-2}$ \\ \hline
$0.2$ & $1.59\times 10^{-3} $ & $1.595\times 10^{-3} $ \\ \hline
$0.1$ & $8.391\times 10^{-5}$ & $8.399\times 10^{-5} $ \\ \hline
$0.08$ & $2.58\times 10^{-5}$ & $2.59\times 10^{-5} $ \\ \hline
$0.05$ & $1.34\times 10^{-6}$ & $1.35\times 10^{-6} $ \\ \hline
$< 0$ & \multicolumn{2}{|c|}{No solutions found.} \\ \hline
  \end{tabular}
\end{table}

The results in Tables 1 and 2, and other similar results for different values of $d$, suggest that
as $\lambda$ approaches the critical value from below, the energy of the lowest bound state
rises and tends towards $0$. This would provide a nice realisation of our expectation from the previous subsection
that there should be no negative energy bound state if $\lambda > \lambda_c$. Thus instability
is according to the criterion (\ref{eq:crit}).

\subsection{Finite cosmological constant}

\subsubsection{Topological black holes}

Let the cosmological constant $c=-L^2$ and let the base manifold
$B$ have negative curvature, $\e=-1$.
The metric is seen to have no cosmological horizon. If $\a=0$ it has
no singularity and an event horizon at $r=\frac{1}{L}$.
These are the so-called topological black holes \cite{van,bir,mann}.
Thus $f$ is
\be\label{eq:ftop}
f(r) = -1 + L^2 r^2.
\end{equation}
If $B$ is a hyperbolic manifold
$B= H^k/\Gamma$, with $\Gamma \subset SO(k,1)$ a suitable discrete
group, then the metric is locally Anti-de Sitter space.

The potential (\ref{eq:V}) is
\be
V(r) = \left[
\frac{d^2-10d+8-4\lambda}{4 r^2}
- \frac{L^2 (d^2+2d)}{4} \right] (1-L^2 r^2) .
\end{equation}
As expected, the potential vanishes on the horizon.
The potential is not necessarily positive outside the event horizon,
so {\it a priori} there
exists the possibility of bound states with negative energy
if $-d^2+10d-8+4\lambda<0$.
The Schr\"odinger equation (\ref{eq:schro}) may be solved exactly in this
case. Two solutions are
\bea\label{eq:solh}
\P_{\pm}(r) & = & r^{(1\pm C)/2} (1-L^2r^2)^{-\w/2L} \times \nonumber \\
& & {\,}_2 F_1 \left(-\frac{\w}{2L}+\frac{1-d\pm C}{4},
-\frac{\w}{2L}+\frac{3+d\pm C}{4};\frac{2\pm C}{2};L^2r^2\right) ,
\eea
where $C = \sqrt{(d-5)^2-4(4+\lambda)}$
and ${\,}_2 F_1 (a,b;c;x)$ is the hypergeometric function.
These are generically linearly independent.

Because there is no cosmological horizon, the perturbation extends
to $r\to\infty$. To consider the asymptotics of (\ref{eq:solh}), use
the following result for hypergeometric functions \cite{spec}
\bea\label{eq:x1x}
{\,}_2 F_1 (a,b;c;x) & = & k_1 (-x)^{-a}
{\,}_2 F_1 \left(a,a-c+1;a-b+1;\frac{1}{x}\right) \nonumber \\
& & + k_2 (-x)^{-b}
{\,}_2 F_1 \left(b,b-c+1;b-a+1;\frac{1}{x}\right) ,
\eea
where\footnote{There is a subtlety here which is that if $d$ is odd
then $a-b$ is a negative integer and one of these gamma functions diverges.
However, the solution (\ref{eq:goodsol}) that we obtain is the solution which
decays at infinity even in these cases.}
\bea
k_1 & = & \frac{\C(c)\C(b-a)}{\C(c-a)\C(b)} , \nonumber \\
k_2 & = & \frac{\C(c)\C(a-b)}{\C(c-b)\C(a)} .
\eea
Using the series expansion of the hypergeometric function about
the origin, this implies that as $x\to\infty$
\bea
{\,}_2 F_1 \left(a,b;c;x\right) & = & k_1 (-x)^{-a}
\left(1 + \frac{a(a-c+1)}{a-b+1}\frac{1}{x} + \cdots\right) \nonumber \\
& + & k_2 (-x)^{-b} \left(1 + \frac{b(b-c+1)}{b-a+1}\frac{1}{x}+\cdots\right) .
\eea
In particular, the power law behaviour at infinity
is $x^{\max(-\Re(a),-\Re(b))}$. In both the solutions of (\ref{eq:solh}) we have
$-\Re(a) > - \Re(b)$, so the $x^{-a}$ term is dominant as $x\to\infty$. It is then easy to check
that the overall leading asymptotic term of both the solutions $\P_{\pm}$ at infinity is
$\mathcal{O}(r^{d/2})$. These solutions are never normalisable in
the sense of (\ref{eq:normal}). However, because infinity is a regular
singular point of this Schr\"odinger equation,
it is possible to take a linear combination of these two solutions that
gives the other allowed power law asymptotics, in this case
$\mathcal{O}(r^{-(2+d)/2})$. This is precisely the linear combination
of $\P_{\pm}$ in which the $x^{-a}$ terms of (\ref{eq:x1x}) cancel. The result is,
using the symmetry ${\,}_2 F_1 (a,b;c;x) = {\,}_2 F_1 (b,a;c;x)$,
\bea\label{eq:goodsol}
\P_3(r) & = & r^{-(2+d)/2+\w/L} (1-L^2 r^2)^{-\w/2L} \times \nonumber \\
& & {\,}_2 F_1 \left(-\frac{\w}{2L}+\frac{3+d+C}{4},
-\frac{\w}{2L}+\frac{3+d-C}{4};\frac{6+2d}{4};\frac{1}{L^2r^2}\right) .
\eea
This solution has acceptable behaviour as $r\to\infty$. We now
need to check the behaviour as $r\to \frac{1}{L}$. To do this we use another
identity of hypergeometric functions \cite{spec}
\bea\label{eq:expr2}
{\,}_2 F_1 (a,b;c;x) & = & h_1 {\,\,}_2 F_1 (a,b;a+b+1-c;1-x) \nonumber \\
& + & h_2 (1-x)^{c-a-b} {\,}_2 F_1 (c-a,c-b;1+c-a-b;1-x) ,  
\eea
where
\bea
h_1 & = & \frac{\C(c)\C(c-a-b)}{\C(c-a)\C(c-b)} , \nonumber \\
h_2 & = & \frac{\C(c)\C(a+b-c)}{\C(a)\C(b)} .
\eea
Applying this to (\ref{eq:goodsol}) we see that generically for the
modes with $\w>0$ that we are looking for, the
leading term as $r\to \frac{1}{L}$ is $\mathcal{O}\left(
(1-L^2r^2)^{-\w/2L} \right)$.
Inserting this and (\ref{eq:ftop}) into the normalisation condition (\ref{eq:normal})
we see that the solution is not square integrable at the horizon.
Furthermore, the solution does not satisfy the boundedness requirement
because $\P$, and hence $\p$, divergs at the horizon.
However, if $C$ is real and $C>3+d$ for some mode, which requires
\be\label{eq:crittop}
\lambda_{min} < -4d ,
\end{equation}
then we may set
\be
\w = \left( \frac{C-3-d}{2}\right) L > 0 .
\end{equation}
For this mode we have that $h_1 = 0$ and $h_2 = 1$ in (\ref{eq:expr2})
because one of the $\C$ functions in the denominator of $h_1$ diverges.
Another way of seing this is that the hypergeometric function is just a
polynomial in this case.
It then follows from (\ref{eq:expr2}) that the mode now has a better behaviour at the horizon,
going as $\mathcal{O}\left(1-L^2r^2)^{\w/2L} \right)$. In particular it
is normalisable and bounded at the horizon. Therefore it gives an instability.
Because the base manifold has negative curvature, the arguments of Appendix A
do not apply in this case and therefore we don't know about the effect of scalar and vector modes.
The statement we can make is that {\it if the base of a massless topological black hole
has a Lichnerowicz spectrum satisfying (\ref{eq:crittop}) then it is unstable}.
Some results on the Lichnerowicz spectrum for negative scalar
curvature Einstein manifolds are collected in \S 5.

\subsubsection{Brane world metric}

Here $f(r)=r^2$. The base manifold is Ricci flat because $\e=0$.
The metric is more familiar in terms of $z=\frac{1}{r}$
\be\label{eq:bw}
ds^2 = \frac{1}{z^2} \Bigl ( dz^2 - dt ^2 + d\tilde{s}^2_d \Bigr ) .
\end{equation}
The $r$ coordinate
extends from the origin to infinity. This metric is of the general brane world
form considered in \cite{rs1, rs2}. More general such metrics will be studied in the section
below on time dependent solutions.

The potential (\ref{eq:V}) is
\be
V(r) = \lambda + \frac{(d^2+2d) r^2}{4} .
\end{equation}
The starred coordinate (\ref{eq:change}) is just $\rs = -z$. The
Schr\"odinger equation (\ref{eq:schro}) may be solved exactly in this
case. The general solution is
\be
\P(r) = A r^{-1/2} I_{(1+d)/2}(\frac{\sqrt{\lambda+\w^2}}{r})
+ B r^{-1/2} K_{(1+d)/2}(\frac{\sqrt{\lambda+\w^2}}{r})  ,
\end{equation}
where $A,B$ are constants and $I_{\nu}, K_{\nu}$ are the modified Bessel
functions. We need to take the real or imaginary part depending on
whether $\sqrt{\lambda+\w^2}$ is real or imaginary.
The term with the $I_{\nu}$ function is always normalisable (\ref{eq:normal}) as $r\to\infty$,
going as $\mathcal{O}(r^{-(2+d)/2})$.
The term with the $K_{\nu}$ function diverges as
$\mathcal{O}(r^{d/2})$
as $r\to\infty$ and hence is never normalisable because $d>1$, although
$\p$ is bounded. We must then check
the $I_{\nu}$ solution as $r\to 0$. In terms of $z$, the normalisability
condition, with an implied real or imaginary part being taken, is
\be
\int_0^{\infty} z I^2_{(1+d)/2}(z\sqrt{\lambda+\w^2}) dz < \infty .
\end{equation}
If $\lambda+\w^2 > 0$ then the integrand diverges exponentially as
$z\to\infty$. There was no chance of a solution in this case because
the energy, $-\w^2$, would have been lower than the minimum of the
potential $\lambda$.
If $\lambda+\w^2 < 0$ then the integrand oscillates with constant
magnitude as $z\to\infty$. In either case, normalisability cannot be
enforced. Furthermore, $\p=\P r^{-\frac{d}{2}}$ is seen to
diverge as $r\to 0$ and so the boundedness condition is not satisfied.
Therefore there are no finite energy solutions and {\it the brane world
metric (\ref{eq:bw}) is {\bf stable} under this perturbation, independent of the
base manifold.}

\subsubsection{Schwarzschild-Tangherlini-Anti-de Sitter black holes}

Set $c = - L^2$ and $\epsilon=+1$. If $\a=0$ then we have a generalised Anti-de Sitter
space that describes the asymptotics of a Schwarzschild-Tangherlini-Anti-de Sitter black hole.
The function $f(r)$ is
\be
f(r) = 1 + L^2 r^2 .
\end{equation}
There are no horizons. The potential is now
\be\label{eq:potads}
V_{\infty}(r) = \left[
\frac{d^2-10d+8+4\lambda}{4 r^2}
+ \frac{(d^2+2d) L^2}{4} \right] (L^2 r^2 + 1).
\end{equation}
Two exact solutions to the Schr\"odinger equation are
\bea
\P_{\pm}(r) & = & r^{(1\pm C^{\prime})/2} (1+L^2r^2)^{-i\w/2L} \times \nonumber \\
& & {\,}_2 F_1 \left(\frac{-i\w}{2L}+\frac{1-d\pm C^{\prime}}{4},
\frac{-i\w}{2L}+\frac{3+d\pm C^{\prime}}{4};\frac{2\pm C^{\prime}}{2};-L^2r^2\right) ,
\eea
with $C^{\prime}=\sqrt{(d-5)^2-4(4-\lambda)}$. This is fairly similar to the topological
black hole case, but we now need to consider different limits. By exactly the same arguments
as for the topological black hole, the solution that is well behaved at infinity, going as
$\mathcal{O}(r^{-(2+d)/2})$, is
\bea\label{eq:gs2}
\P_3(r) & = & r^{-(2+d)/2+i\w/L} (1+L^2 r^2)^{-i\w/2L} \times \nonumber \\
& & {\,}_2 F_1 \left(\frac{-i\w}{2L}+\frac{3+d+C^{\prime}}{4},
\frac{-i\w}{2L}+\frac{3+d-C^{\prime}}{4};\frac{6+2d}{4};\frac{-1}{L^2r^2}\right) .
\eea
By using (\ref{eq:x1x}) again, we see that the behaviour as $r\to 0$ is
$\mathcal{O}\left(r^{(1-C^{\prime})/2}\right)$. This both converges and is normalisable if
$d^2-10d+8+4\lambda_{min} < 0 $
which is unsurprisingly also the condition for the
potential (\ref{eq:potads}) to be negative and indeed unbounded below. There is a continuum of
negative energy bound states with $\w > 0$ in this case.
None of these solutions satisfy the boundedness condition because $\p=\P r^{-\frac{d}{2}}$
always diverges at the origin. Thus the massless case is stable against the perturbation.
The phenomenon of the hypergeoemtric series terminating for special values of $\w$ to
give a well behaved mode does not occur here because of the $i$ in front of the $\w$ in
\ref{eq:gs2}. More interestingly, if
\be
\lambda_{min} < 4 - \frac{(5-d)^2}{4} ,
\end{equation}
then the solution is oscillatory in the inner regions.
The oscillatory behaviour suggests we can make a statement about massive black holes also.
The condition is clearly
the same (\ref{eq:crit}) as we found before in the case of vanishing cosmological constant.
This is perhaps not surprising given that the potentials, (\ref{eq:vinf}) and (\ref{eq:potads}),
are the same near the origin. The comments of \S 3.2 should go
through. However, there is now a second lengthscale in the problem, $1/L$, which could
delay the onset of the oscillations until beyond the event horizon, in which case there will
be no solution and no instability.

A few things may be said more concretely. In the massive case
\be
f(r) = 1 - \left(\frac{\a}{r}\right)^{d-1} + L^2 r^2 .
\end{equation}
The potential becomes
\be
V(r) = V_{\infty}(r) +\frac{1}{\a^2} \left(\frac{\a}{r}\right)^{d+1} \left[ \frac{10d-8-4\lambda-2d L^2 r^2}{4}
-\left(\frac{\a}{r}\right)^{d-1} \frac{d^2}{4} \right].
\end{equation}
The potential has the expected property that it vanishes at the horizon, where $f(r)=0$. Furthermore,
when $d^2-10d+8+4\lambda > 0 $ it is everywhere positive,
as was the asymptotic (massless) potential (\ref{eq:potads}). Thus there will be no instability in
these cases. We now need to see numerically when a solution exists and what the role of the
new lengthscale is.

\subsubsection{Numerical results for Schwarzschild-Tangherlini-Anti-de Sitter}

We wish to use the methods of \S 3.2.2 to examine the effect of the
new lengthscale $1/L$. First note that the Schr\"odinger equation is
now invariant under the scaling: $\a\to k\a, r\to kr, \w\to\w/k,
L\to L/k$. Previously, when there was no $L$ we used this to set
$\a=1$, which was the location of the horizon. Again, we want to scale
the horizon to $1$. This will now require scaling so that $\a =
(1+L^2)^{1/(d-1)}$. There is a scale-invariant dimensionless mass
\be
M = \a L ,
\end{equation}
which allows us to talk about large and small black holes
independently of the scaling used. We expect the criterion for
instability to be the same as for Schwarzschild black holes
(\ref{eq:crit}) when the AdS black hole is small $M \to 0$. As we
increase $M$ we expect the black hole to be stabilised by the
cosmological constant. This behaviour indeed happens and is
illustrated in table 3. The numerics were done as in \S 3.2.2
For the case $d=4$ (hence spacetime dimension
$D=6$) and with $\lambda$ satisfying the instability criterion by
$\lambda = \lambda_c-1$ we see that the unstable mode is stabilised
if $M > 0.058...$.

\begin{table}[h]
{\bf Table 3: } Lowest bound states for $\lambda=\lambda_c-1, d=4$ ($D=6$)
generalised Schwarzschild-AdS. \\
  \begin{tabular}{|c|c|c|c|} \hline
$L$ & $M = \a L$ & Lower bound for $\w_{max}$ & Upper bound for $\w_{max}$ \\ \hline \hline
$0$ & $0$ & $0.1331$ & $0.1335$ \\ \hline
$0.01$ & $0.0100003...$ & $0.131$ & $0.132$ \\ \hline
$0.05$ & $0.04004...$ & $0.0591$ & $0.00599$ \\ \hline
$0.055$ & $0.05505...$ & $0.0264$ & $0.0269$ \\ \hline
$0.057$ & $0.05706...$ & $0.0005$ & $0.00051$ \\ \hline
$> 0.058$ & $0.05806...$ & \multicolumn{2}{|c|}{No solutions found.} \\ \hline
  \end{tabular}
\end{table}

This will be the generic behaviour. {\it A given base space will have a
minimum Lichnerowicz eigenvalue, $\lambda$. If this is less than the
critical value $\lambda_c$ then we can find a critical value for the
dimensionless mass, $M_c$, such that if $M_c < M$ then the unstable
mode is stabilised. If $M_c > M$ then the AdS black hole is unstable}.
Alternatively, we could think of the mass as altering
the expression for $\lambda_c$.

\subsubsection{Schwarzschild-Tangherlini-de Sitter black holes}

Set $c = L^2$ and $\epsilon=+1$. This gives us a generalised
Schwarzschild-Tangherlini-de Sitter black hole. We have,
\be
f(r) = 1 - \left(\frac{\a}{r}\right)^{d-1} - L^2 r^2 .
\end{equation}
There is a
cosmological horizon at finite radius, and therefore we cannot discuss
an asymptotic solution to the Schr\"odinger equation, because the mass
term is not negligible near the horizon.
We have all the information necessary to tackle this problem numerically, but various
cases must be considered separately depending on the values of $\a$
and $L$. This is somewhat out of the main line of development of this
work and so it will not be considered here.

\section{Time-dependent solutions}

\subsection{Generalised de Sitter}

The metric form (\ref{eq:metric}) also covers a range of cosmological solutions. For
example, a generalised de Sitter metric may be written
\be
ds^2 = \frac{-dr^2}{L^2 r^2-1} + r^2 d\tilde{s}^2_d ,
\end{equation}
where $r$ is the time coordinate now and $L$ is a constant.
This is of the form (\ref{eq:metric}) with $f=0$.
Note that $d$ would now be 3, not 2, in the usual four dimensional case. Consider
a perturbation, using the same notation as in (\ref{eq:mode}),
\be\label{eq:tpert}
h_{\a\b}(x) = \widetilde{h}_{\a\b}(\tilde{x}) r^2 \p(r) ,
\end{equation}
and impose the Einstein equation, $\d R_{\a\b}=d L^2 h_{\a\b}$. There is now
a cosmological term which is that of ordinary de Sitter space in $D=d+1$ dimensions.
The equation for the perturbation $\p$ is most familiar in terms of the coordinates
\be
\cosh Lt = Lr ,
\end{equation}
where the metric becomes
\be
ds^2 = -dt^2 + \frac{\cosh^2 Lt}{L^2} d\tilde{s}^2_d ,
\end{equation}
and the equation (\ref{eq:dl}) becomes the equation for a scalar field on de Sitter space
\be
\frac{d^2\p}{dt^2} + d L \tanh Lt \frac{d\p}{dt} + \frac{L^2}{\cosh^2 Lt} (\lambda+2-2d) \p = 0 .
\end{equation}
This provides a check on our expression (\ref{eq:dl}) and also shows that
at late times as $t\to\infty$ the leading term in each of the two linearly independent
solutions is
\be
\p \sim A + B e^{- d L t} .
\end{equation}
With $A,B$ constants.
{\it Thus perturbations are frozen in, independent of the dimension and the form of the
base Einstein manifold $B$.} This is just the behaviour of such perturbations
in standard four dimensional inflationary metrics \cite{bg}.

\subsection{Generalised Anti-de Sitter}

Generalised Anti-de Sitter space can be treated similarly. Write the
metric as
\be
ds^2 = \frac{-dr^2}{-L^2 r^2+1} + r^2 d\tilde{s}^2_d ,
\end{equation}
where again $r$ is the time coordinate and $L$ a constant. The base
manifold $B$ must now have negative curvature and would be $H^d$ for
Anti-de Sitter itself. Consider the perturbation
\be
h_{\a\b}(x) = \widetilde{h}_{\a\b}(\tilde{x}) r^2 \p(r) ,
\end{equation}
and impose the Einstein equation, $\d R_{\a\b}=-d L^2 h_{\a\b}$. This
is as for the de Sitter case considered previously
but with a negative cosmological constant. The
familiar coordinates for the space are
\be
\sin Lt = Lr .
\end{equation}
The metric in these coordinates is
\be
ds^2 = -dt^2 + \frac{\sin^2 Lt}{L^2} d\tilde{s}^2_d .
\end{equation}
These coordinates make explicit the Big Bang and Big Crunch
singularities at $t=0$ and $t=\frac{\pi}{L}$, unless the spacetime
is Anti-de Sitter. The equation for the perturbation becomes
\be\label{eq:eqads}
\frac{d^2\p}{dt^2} + d L \cot Lt \frac{d\p}{dt} + \frac{L^2}{\sin^2 Lt} (\lambda+2d-2) \p = 0 .
\end{equation}
The general solution to this equation is
\bea\label{eq:solads}
\p & = & A \left( \sin Lt \right)^{(1-d)/2} P_{(d-1)/2}{}^{C/2}(\cos
Lt) \nonumber \\
& + & B \left( \sin Lt \right)^{(1-d)/2} Q_{(d-1)/2}{}^{C/2}(\cos Lt) ,
\eea
where $A,B$ are constants and as before $C =
\sqrt{(d-5)^2-4(4+\lambda)}$. $P_{\nu}{}^{\mu}$ and
$Q_{\nu}{}^{\mu}$ are Legendre functions of the first and second kind,
respectively. These may be expressed in terms of hypergeometric
functions as follows \cite{ho}
\bea\label{eq:leg}
P_{\nu}{}^{\mu}(x) & \propto &
(x^2-1)^{\mu/2} x^{\nu-\mu}
{\,}_2 F_1\left(\frac{\mu-\nu}{2},\frac{\mu-\nu+1}{2};\frac{1}{2}-\nu;\frac{1}{x^2}\right) , \nonumber \\
Q_{\nu}{}^{\mu}(x) & \propto &
(x^2-1)^{\mu/2} x^{-\nu-\mu-1}
{\,}_2 F_1\left(\frac{\nu+\mu+1}{2},\frac{\nu+\mu+2}{2};\nu+\frac{3}{2};\frac{1}{x^2}\right) .
\eea
In the present case $x=\cos Lt$ with $0 \leq t \leq
\frac{\pi}{L}$. Thus we need to check regularity properties at
$t=0,\frac{\pi}{2L},\frac{\pi}{L}$, corresponding to $x=1,0,-1$.

It is easy to check using (\ref{eq:x1x}) that both solutions are
finite at $t=\frac{\pi}{2L}$, that is $x=0$. Further, it is clear from
(\ref{eq:leg}) and (\ref{eq:solads}) that behaviour as $t\to\frac{\pi}{L}$
will be the same as for $t\to 0$, up to phases in front of
each Legendre function. In particular this means that regularity
properties will be the same at these points. Both solutions of
(\ref{eq:solads}) diverge as $\mathcal{O}(t^{(1-d-C)/2})$ as
$t\to 0$. Therefore they also diverge as $t\to\frac{\pi}{L}$. However,
$t=0$ is a regular singular point of the equation (\ref{eq:eqads}) and
therefore there will be a linear combination of these solutions that
has the other allowed power law behaviour as $t\to 0$, namely
$\mathcal{O}(t^{(1-d+C)/2})$. This will be oscillatory and divergent
if $C$ is pure imaginary. If $C$ is real it will converge if
$C \geq d-1$ and diverge otherwise. There is also the possibility that
well behaved modes will exist for special values of $C$ where the
hypergeometric function becomes a polynomial.

The main conculsion of the previous paragraph is that there are always
modes that if excited at some finite time will diverge in the
future. {\it Thus this AdS cosmology is unstable, independent of the base
manifold and the dimension.}

\subsection{Ricci flat Lorentzian base (Brane world metrics II)}

Let the base $B$ be a $d$-dimensional Ricci flat spacetime. Suppose we
know the spectrum of Lichnerowicz modes on $B$ with
eigenvalues $\lambda$ and such that the modes grow
in time. In particular, if this spectrum includes a zero mode, then the
spacetime $B$ is unstable.

The spacetime $B$ may be embedded in a $D=d+1$ dimensional Einstein
manifold with negative scalar curvature
\be\label{eq:ds2}
ds^2 = \frac{1}{z^2} \Bigl ( dz^2 + d\tilde{s}^2_d \Bigr )
= \frac{dr^2}{r^2}+r^2 d\tilde{s}^2_d ,
\end{equation}
For example, if $B$ is a black hole metric we obtain a
black string in AdS spacetime.
The change of variables in (\ref{eq:ds2}) is of course
$r=1/z$. We would like to see whether
this spacetime is unstable under any of the growing modes in $B$. Consider
the perturbation
\be
h_{\a\b}(x) = r^2 \p(r) \tilde{h}_{\a\b}(\tilde{x}) ,
\end{equation}
where $\tilde{h}_{\a\b}$ is a Lichnerowicz eigenmode on the base with eigenvalue
$\lambda$. From (\ref{eq:dl}), ignoring the terms with $f$s, the equation for $\p(r)$ coming
from $\D_L h_{\a\b} = -2d h_{\a\b}$ is
\be\label{eq:sl2}
\frac{d^2\p}{dr^2} + \frac{d+1}{r} \frac{d\p}{dr} - \frac{\lambda \p}{r^4} = 0 .
\end{equation}
The general solution to this equation for $\lambda > 0$ is
\be
\p(r) = A r^{-d/2} I_{d/2}\left(\frac{\lambda^{1/2}}{r} \right)
+ B r^{-d/2} K_{d/2}\left(\frac{\lambda^{1/2}}{r} \right) ,
\end{equation}
where $A,B$ are constants and $I_{\mu},K_{\mu}$ are the modified Bessel functions. If
$\lambda < 0$ then the expression is most transparent if we let $\lambda\to -\lambda$
and replace $I_{\mu},K_{\mu}$
by the Bessel functions $J_{\mu},Y_{\mu}$. Finally, if $\lambda = 0$ then the solution is
\be\label{eq:lis0}
\p(r) = A + B r^{-d} .
\end{equation}

To see which, if any, of these solutions are acceptable we need to find the energy of the
perturbations. Assuming that $B$ has a timelike Killing vector, 
this is similar to the argument in \S 3.1
\bea\label{eq:enerlor}
E & \propto & \int t^{\mu \nu} n_{\mu}\xi_{\nu}\sqrt{g^{(d)}} d^{d}x =
\int t^{\mu \nu} n_{\mu}\xi_{\nu} \sqrt{\tilde{g}^{(d-1)}} r^{d-2} dr d^{d-1}\tilde{x} \nonumber \\
& \sim & \int \p^2 r^{d-3} dr ,
\eea
where we used the metric (\ref{eq:ds2}) and $n_0 \propto r$, $\xi_0 \propto r^2$,
$t_{00}\sim \p(r)^2$. Note that $\tilde{g}^{(d-1)}$ is the spatial metric on $B$.
It is easy to see that the normalisation condition (\ref{eq:enerlor}) is the same as
the norm of the equation (\ref{eq:sl2}) cast in Sturm-Liouville form. The eigenvalue
of the Sturm-Liouville problem is now $\lambda$ and the weight function
is seen to be $r^{d-3}$, consistent with (\ref{eq:enerlor}).
The boundedness condition is this context is simply that $\p(r)$ remains finite in
the range $0 \leq r < \infty$.

If there is a $\lambda=0$ mode, the $d$ dimensional spacetime $B$ is unstable.
However, from (\ref{eq:lis0})
it is clear that none of these solutions are normalisable in the $d+1$ dimensional
spacetime. Further, all except the constant solutions are not bounded.
Therefore the unstable mode does not carry over to the full spacetime
because it no longer has finite energy.

For $\lambda > 0$ we see that the term with the $I_{\mu}$ Bessel function has an
exponential divergence as $r\to 0$ whilst the term with the $K_{\mu}$ Bessel
function goes to zero exponentially and hence is normalisable at the origin. However, this term goes as
$\mathcal{O}(r^0)$ as $r\to\infty$ and hence, although bounded, is not
normalisable at infinity because $d\geq 2$. Thus there are no normalisable
solutions with $\lambda > 0$.

For $\lambda < 0$, as $r\to\infty$, the term
with the $Y_{\mu}$ Bessel function goes as $\mathcal{O}(r^0)$,
and therefore is not
normalisable at inifinity. The term with the $J_{\mu}$ Bessel function goes
as $\mathcal{O}(r^{-d})$ and therefore is normalisable. 
As $r\to 0$, the $J_{\mu}$ term oscillates as
$\mathcal{O}(r^{(1-d)/2}\cos\frac{(-\lambda)^{1/2}}{r})$ and hence are
not normalisable (\ref{eq:enerlor}) or bounded as $r\to 0$. 
Thus there are no normalisable modes of this type.

In conclusion, embedding a Ricci flat spacetime $B$ into a higher
dimensional spacetime with cosmological constant as in (\ref{eq:ds2})
will stabilise at the linear level any unstable modes of $B$. Furthermore, no
other unstable modes of the form we consider appear. This should be
contrasted with a similar embedding into a higher dimensional
Ricci flat spacetime where the stability properties get worse due
to negative Lichnerowicz modes in the initial spacetime \cite{real},
such as the Gregory-Laflamme instability of nonextremal black strings
\cite{gl1,gl2,gl3}. It was argued in \cite{laf} that a Gregory-Laflamme
instability existed also for black strings in AdS spacetime. However, the
perturbed mode presented there, which agrees as a special case
with the modes we have just considered, did not have $h^{\a}{}_{\b}$ bounded
as is required in a linearised analysis. The phenomenon of perturbations to
brane world metrics diverging in the bulk has been observed before \cite{cg} and is
related to the bad behaviour of the curvature at the horizon.
Bounded modes, and hence the instability, could reappear if one modifies
the setup, such as by adding a negative tension brane at finite position.

One might worry that the instability of a very thin short black string
should not be affected by immersion in an Anti-de Sitter spacetime
with large radius. This may well be true. However, if one assumes that
the black string runs all the way to the horizon there seems to be no
way of avoiding our conclusions, although there is already a
singularity near the horizon in the unperturbed metric \cite{chr}. The
methods used here can say very little about what would happen for a
``cigar-like'' black string configuration.

\subsection{Double analytic continuation}

Higher dimensional versions of the Schwarzschild bubble solution \cite{wit}
have been considered recently in a search for well-behaved time dependent backgrounds
in which to study string theory \cite{afhs}. Ultimately the Kerr solutions turn out to be
more appropriate, but the Schwarzschild case contains many of the relevant features.
The solution, obtained via double
analytic continuation of the Schwarzschild-Tangherlini solution is
\be\label{eq:bubble}
ds^2 = \left[ 1 - \left( \frac{\a}{r} \right)^{d-1} \right] d\psi^2 +
\frac{dr^2}{1 - \left( \frac{\a}{r} \right)^{d-1}} + r^2 (-d\tau^2 + \cosh^2\tau d\Omega^2_{d-1}),
\end{equation}
where $r\geq \a$, $\psi$ has period $\frac{4\pi\a}{d-1}$ and $d\Omega^2_{d-1}$
is the round metric on $S^{d-1}$.  The coordinate $\psi=it$ is the Wick rotated
time from the black hole solution and $i\tau=\theta-\frac{\pi}{2}$, where $\theta$
was the usual angular parameter on $S^d$ and $\tau$ is now the time on
$dS_d$. Recently, such ``bubbles of nothing'' have also been
considered by analytically continuing AdS-Schwarzschild black holes \cite{br1,br2}.

A classically stable background is needed for string theory.
Classical instabilities are manifested in the string worldsheet theory as a
renormalisation group flow arising from higher string loops with a fixed
point that is not close to the original background in string units.

It was argued in \cite{afhs} that
classical stability of (\ref{eq:bubble}) follows from the classical stability of
the corresponding black hole solution. Any mode on the bubble
must be periodic in the $\psi$ variable. But $e^{-i\w\psi} = e^{wt}$,
so these correspond to growing modes of the black hole with certain
frequencies $\w$ allowed by the periodicity of $\psi$. Furthermore, the part
of the black hole mode that is a harmonic on $S^d$ becomes a harmonic
on $dS_d$ with the same
eigenvalue. In the four dimensional case of \cite{rw,vish,zer} these modes are scalar
and vector harmonics, whilst in our case we are considering tensor harmonics \cite{hig}.
The equation for the radial dependence is the same in the black hole and bubble cases
and therefore the criterion for the existence of solutions
is the same. On the sphere, the harmonics are trigonometric functions of $\theta$
and these become hyperbolic functions of $\tau$, which correspond to growing
and hence unstable modes on the bubble.

The conclusion of the previous paragraph is that any unstable mode on the
black hole with appropriate $\w$ signals an unstable mode on the bubble. Conversely,
any mode on the bubble signals an unstable mode on the black hole.
Thus we have extended the stability arguments of the higher dimensional Schwarzschild bubble in
\cite{afhs} to include the tensor mode that is being considered throughout this paper.

There is a subtlety, however. After doing
the double analytic continuation, we must redo the calculations of energy because
the timelike Killing vector has changed, if indeed there is still a timelike Killing vector at all.
The calculation is very similar to the Lorentzian base case (\ref{eq:enerlor}). There
will typically be horizons on the base manifold across which the Killing vector changes sign,
as in the de Sitter base of (\ref{eq:bubble}). In this case, the integration over a spacelike
hypersurface of the base should be restricted to the region inside the horizon. The integration
over $r$ is not changed.
\bea\label{eq:ebub}
E & \propto & \int t^{\mu \nu} n_{\mu}\xi_{\nu}\sqrt{g^{(d+1)}} d^{d+1}x \sim
\int t^{00} n_{0}\xi_{0} \sqrt{\tilde{g}^{(d-1)}} r^{d-1} dr d\psi d^{d-1}\tilde{x} \nonumber \\
& \sim & \int \p^2 r^{d-2} dr = \int \P^2 r^{-2} dr ,
\eea
which is indeed different to the energy of the corresponding black hole perturbation (\ref{eq:ener}).
As a check of this expression, we can see that this is the same normalisability condition that we get from the
Sturm-Liouville problem for $\p$. The Sturm-Liouville equation remains (\ref{eq:sl}), but
the eigenvalue associated with time evolution is no longer $\w$, but rather $\lambda$, as was
commented for the case of the Lorentzian base.
The weight function is thus seen to be $r^{d-2}$ rather than the $r^d/f$ that we had for the
black hole. This is in agreement with (\ref{eq:ebub}). The condition for boundedness
remains the same, $\p$ must not diverge. We may now go through the results from the black hole spacetimes
of \S 3 again and recheck for finite energy using (\ref{eq:ebub}). We see that the
existence and nonexistence of finite energy solutions remains the same in each case.

If the Euclidean metric on the base manifold $B$ of the generalised
black hole solutions, (\ref{eq:metric}) and (\ref{eq:fform}),
admits an analytic continuation to a Lorentzian metric, then one may construct a generalised
``bubble of nothing'' with metric
\be\label{eq:genbub}
ds^2 = f(r) d\psi^2 +
\frac{dr^2}{f(r)} + r^2 d\tilde{s}^2_{d},
\end{equation}
with the notation of (\ref{eq:bubble}) and now $d\tilde{s}^2_{d}$ is
a Lorentzian metric.

Assuming that the tensor harmonics on the Lorentzian base manifold
are growing modes in time, the argument for the $S^d$ above case goes through
also in the generalised case. Stability is thus related to the stability
of a generalised black hole solution.
We could also see this directly from the relationship
between the Lichnerowicz operator on the full manifold and on the base
(\ref{eq:dl}): perturbations of (\ref{eq:genbub}) that are periodic in
$\psi$ will give the same radial equation as perturbations of the generalised
black hole metrics with exponential growth in $t$. Thus, for example,
in the generalised Schwarzschild-Tangherlini case,
the criterion (\ref{eq:crit}) will be the same. However, there is an important caveat: 
the frequencies that cause instabilities for the generalised black hole will only
cause an instability in the bubble spacetime if they respect the periodicity of
$\psi$, that is
\be
\w = \frac{N(d-1)}{2\a} ,\quad\quad N \in \mathbb{Z} .
\end{equation}
In practice, this is unlikely to be true for any of the bound states in the discrete
spectrum. Therefore we expect that the bubble spacetimes will generally be stable against the perturbation
even when they violate the criterion (\ref{eq:crit}).

\section{Spectra of the Lichnerowicz operator}

This section collects and applies mostly known results about the spectrum
of the Lichnerowicz operator on various Einstein manifolds. In our
considerations of the static metrics
above we found that in the case of Schwarzschild-Tangherlini,
Schwarzschild-Tangherlini-Anti-de Sitter and topological black holes,
stability of the spacetime against the perturbation (\ref{eq:gauge}),
(\ref{eq:extra}) depends on the spectrum of the Lichernowicz operator
on the base manifold $B$ (\ref{eq:crit}). The objective of this section is to check
that the cases we expect to be stable, i.e. round spheres, are stable and
also to find examples of spaces that are not stable, thus showing that
the criterion for instability is not vacuous. It would be nice to have more spectra at
hand, in particular for Einstein manifolds that are topological spheres such as
the Bohm metrics \cite{bohm}. We discuss first Einstein manifolds of
positive scalar curvature, relevant for the Schwarzschild and
Schwarzchild-AdS cases. We then discuss manifolds of negative scalar
curvature, which are relevant for the topological black holes.

In this section, tildes will be omitted from the metric perturbations on the base manifold
$B$ and we will
consider transverse tracefree perturbations
\be
g_{\a\b} \to g_{\a\b} + h_{\a\b}, \quad\quad h^{\a}{}_{\a}
= \nabla^{\a} h_{\a\b} = 0 .
\end{equation}
Define $\Lambda$ by $R_{\a\b} = \Lambda g_{\a\b}$.

\subsection{Round spheres}

When the base manifold is a unit sphere, Asymptotically Conical (AC)
just means Asymptotically Flat (AF).
The unit sphere $B = S^d$ has curvature
\be
R_{\a\b\c\d} = g_{\a\c} g_{\b\d} - g_{\a\d} g_{\b\c} ,
\end{equation}
and therefore the Lichnerowicz operator is
\be
(\D_L h)_{\a\b} = (-\nabla^2 h)_{\a\b} + 2d h_{\a\b} .
\end{equation}
However, the Laplacian on the sphere is a positive operator and therefore
$\lambda_{\min} \geq 2d$. In fact, the spectrum of transverse tracefree
tensor harmonics on $S^d$, $d \geq 3$, is known to be \cite{ro}
\be
(-\nabla^2 h)_{\a\b} = \left[ k(k+d-1)-2 \right] h_{\a\b} .
\end{equation}
But $2d$ is larger than the critical eigenvalue of
(\ref{eq:crit}) for all $d$, and therefore the spacetime is stable against
this perturbation. In particular this means that the AF Schwarzschild-Tangherlini
black holes are stable to the perturbation, as we should expect. The
spectrum of quotients of the sphere by a finite group has the same
lower bound and so the resulting ALF spaces also give stable black holes.

\subsection{Product metrics}

Let the base be metrically a product of Einstein manifolds.
It is well known \cite{dnp1,dfghm} that there is a Lichnerowicz transverse
tracefree zero mode in which one component expands and the other shrinks,
keeping the total volume constant. Specifically, if the metric decomposes as
\be
d\tilde{s}^2_d = ds^2_{n} + ds^2_{d-n} ,
\end{equation}
then the mode
\be
h_{\a\b} = \left(
\begin{array}{cc}
\frac{\e}{n} g_n & \\
 & \frac{-\e}{d-n} g_{d-n} \\
\end{array}
\right)_{\a\b} ,
\end{equation}
is easily seen to be tracefree, transverse and with zero Lichnerowicz eigenvalue.
Comparing with (\ref{eq:crit}), this implies that the spactime is unstable
with a product base for $d<9$. It is intriguing that this is the same critical
dimension for instability as was found in the context of generalised Freund-Rubin
compactifications to $AdS\times B_d$ \cite{dfghm}, where the general
dependence on the dimension is different.

\subsection{Five and seven dimensional base}

In five dimensions, one has the Einstein manifolds $T^{pq}$ which are
$U(1)$ bundles over $S^2\times S^2$. See e.g. \cite{gm} for the metric and
other details. We will not make their choice
of normalisation $a=1$ and $a$ will appear as a parameter related to the
overall scale of the metric. Instead we require that $\Lambda=4$, as this
is the normalisation of the unit sphere in five dimensions. We have \cite{gm}
\be
\left(\frac{q}{p}\right)^2 = \frac{1-8a^2}{(4 a^2-1)(12 a^2-1)^2} ,
\end{equation}
which is seen to imply
\be\label{eq:range}
\frac{1}{8} \leq a^2 \leq \frac{1}{4} .
\end{equation}
The minimum Lichnerowicz eigenvalue is
\be
\lambda_{min} = \left[
 12a^2 - \sqrt{784a^4-240a^2+20} \right] \frac{1}{a^2}.
\end{equation}
The critical value in four dimensions (\ref{eq:crit}) is $\lambda_c = 4$.
For the range of $a^2$ in (\ref{eq:range}), we see that
$\lambda_{min} \leq \lambda_c$, with equality ocurring for the $T^{11}$
case where $a^2 = \frac{1}{6}$. Thus the $T^{11}$ black hole is 
stable whilst all the
other $T^{pq}$ black holes are unstable.
This is precisely the behaviour that was observed in the context
of Freund-Rubin compactifications in \cite{gm}.

In seven dimensions, the Einstein manifolds $M^{pqr}$, which are
$U(1)$ bundles over $\mathbb{CP}^2\times S^2$, and $Q^{n_1 n_2 n_3}$,
which are $U(1)$ bundles over $S^2\times S^2\times S^2$,
have been studied in the context of Freund-Rubin compactifications \cite{pp1,pp2}.
Again, results will be quoted. Interestingly enough, in seven dimensions the manifolds with
the required normalisation, $\Lambda=6$, turn out to be stable precisely when the corresponding
Freund-Rubin supergravity compactifications are stable. Thus the bounds we obtain are familiar.

For $M^{pqr}$, the stability depends on $\frac{p}{q}$ and $\Lambda$ as follows
\be
\lambda_{min} = \Lambda\left[\frac{4+4x-2(25-48x+32x^2)^{\frac{1}{2}}}{1+2x} \right] ,
\end{equation}
where $x$ is defined by
\be
\left(\frac{p}{q} \right)^2 = \frac{2x-1}{x^2 (3-2x)} .
\end{equation}
This implies $\frac{1}{2} \leq x \leq \frac{3}{2}$. We are interested in the
case $\Lambda = 6$. Given that the critical value in
seven dimensions is $\lambda_c = 3$ (\ref{eq:crit}), we have that the solution will
be stable to this perturbation for
\be
\frac{9}{14} \approx 0.64\ldots < x <
1.15\ldots \approx \frac{39}{34},
\end{equation}
and unstable for $x$ outside this interval. Translated into an interval for
$\frac{p}{q}$ this becomes
\be
\frac{7\sqrt{6}}{27} \approx 0.63\ldots < \left| \frac{p}{q} \right| <
1.18\ldots \approx \frac{17\sqrt{66}}{117} .
\end{equation}
Thus for this range of $p,q$ we have solutions that are stable against
perturbations of the kind considered in this work, and therefore stable
overall in the cases of appendix A.
This means that the Schwarzchild-Tangherlini
solution with the sphere as base space is not the only classically stable solution of the
form (\ref{eq:metric}), contrary perhaps to initial expectations. These spaces
are not necessarily simply connected, having fundamental group $\mathbb{Z}_r$.

The situation is more complicated for $Q^{n_1 n_2 n_3}$. We have
$\lambda_{min} = \c_{min} \Lambda$, where $\c_{min}$ is the smallest root of
\be
\c^3 - 6\c^2 + 20(\a_1\a_2+\a_2\a_3+\a_3\a_1)\c - 56\a_1\a_2 \a_3 = 0 ,
\end{equation}
with the $\a_i$s defined by
\bea
& & \left( \frac{n_1}{n_2} \right)^2 = \frac{\a_1 (1+\a_2)^2}{\a_2 (1+\a_1)^2}
+ \mbox{ cylic} \nonumber \\
& & \a_1 + \a_2 + \a_3 = 1 .
\eea
We will look at the simplified case in which $\a_2 = \a_3$, implying $n_2 = n_3$.
Further, set $\Lambda=6$. The resulting range of stability to this perturbation
is found to be \cite{pp2}
\be
\frac{3 \sqrt{2}}{5} \approx 0.85\ldots < \left| \frac{n_2}{n_1} \right| <
1.18\ldots \approx \frac{17 \sqrt{66}}{177}
\end{equation}
Again we have found a countable infinity of AC Schwarzschild-Tangherlini
solutions that are stable to this perturbation. The upper bound is the same as previously.
The fundamental group is $\mathbb{Z}_k$, where $k$ is the greatest common divisor
of $n_1,n_2,n_3$.

Finally, because the bound for stability on the Lichnerowicz operator is the same for a seven
dimensional base as the Freund-Rubin bound, all the manifolds that are supersymmetric in the
supergravity context give a stable spacetime. A list of such spaces can be found in \cite{dnp};
examples are the squashed seven sphere and $SO(5)/SO(3)_{max}$. 

\subsection{Negative scalar curvature manifolds}

Hyperbolic space and its quotients by appropriate finite groups have curvature
\be
R_{\a\b\c\d} = - g_{\a\c} g_{\b\d} + g_{\a\d} g_{\b\c} ,
\end{equation}
and therefore the Lichnerowicz operator is
\be
(\D_L h)_{\a\b} = (-\nabla^2 h)_{\a\b} - 2d h_{\a\b} .
\end{equation}
The rough Laplacian is a positive definite operator on normalisable
modes and therefore the Lichnerowicz spectrum is bounded below by
$-2d$. Comparing with (\ref{eq:crittop}) we see that the corresponding
topological black holes are always stable against the tensor
perturbations.

The spectra of negative scalar curvature Einstein-K\"ahler manifolds
was studied in \cite{fried} with the result that for these manifolds
the Lichnerowicz operator on transverse tracefree tensors is bounded
below by $2-2d$. Therefore these manifolds always give black holes that are
stable against such perturbations.

\section{Conclusions and open issues}

Let us summarise the results of this paper.

\vspace{0.5cm}

\noindent {\bf (a)} Spacetimes of the form (\ref{eq:metric}) admit
perturbations that are transverse tracefree tensors on the base
Einstein manifold. The stability of these spacetimes against such
perturbations depends on the spectrum of the Lichnerowicz operator on
the base manifold.

\vspace{0.5cm}

\noindent {\bf (b)} Higher dimensional generalised Schwarzschild black
holes are stable if and only if the Lichnerowicz spectrum on the base
manifold is bounded below by the critical value of
(\ref{eq:crit}). This statement includes all perturbations to the
spacetime. The same criterion holds for Schwarzschild-AdS black holes,
although in this case large black holes become stabilised. Examples of
base manifolds leading to stable and unstable black holes are given in
\S 5.

\vspace{0.5cm}

\noindent {\bf (c)} (Generalised) Topological black holes have a
criterion for instability due to tensor modes on the base
(\ref{eq:crittop}). An Einstein-K\"ahler base always gives a stable
black hole, as do quotients of hyperbolic space. The
brane world metrics of \S 3.3.2 are always stable against these tensor
perturbations. Time-dependent tensor pertubations in generalised de
Sitter spacetime, considered as a cosmological spacetime, are always frozen in
whilst in generalised Anti-de Sitter spacetime they always include an
unstable mode.

\vspace{0.5cm}

\noindent {\bf (d)} A spacetime with no cosmological constant
may be embedded in a higher dimensional spacetime with a negative
cosmological constant. Lichnerowicz zero modes which correspond to instabilities of
the initial spacetime become stabilised. At the linearised level, no new instabilities
of the form we consider are introduced.
In particular this means that the perturbative Gregory-Laflamme instability does not occur for
AdS black strings in five dimensions, contrary to previous claims
\cite{laf}. See discussion in \S 4.3.

\vspace{0.5cm}

\noindent {\bf (e)} Double analytic continuation of generalised black
hole spacetimes, if admissable, produces a generalised ``bubble of
nothing'' spacetime. These spacetimes are generically more stable to the
analytically continued tensor perturbation than the corresponding
black hole. However, one needs to redo energy calculations because the
time direction has changed.

\vspace{0.5cm}

Some issues remain to be addressed. For the perturbation analysis,
these include a numerical study of the stability
of generalised de Sitter black holes, an explicit calculation of the
vector and scalar modes to confirm the arguments of appendix A and to
extend them to cases where the base manifold has negative or zero
curvature and an extension of the arguments to non-vacuum solutions.

Regarding the Lichnerowicz spectrum, the spectrum is not known for
many interesting metrics, such as the Bohm metrics.
It is possible that the methods of the present work may be used to study aspects of
this spectrum.

In other directions, it might be interesting to see whether rotating
black holes admit a similar generalisation. Also, to see whether these
instabilities are useful in the context of gravity-gauge theory
dualities.

\vspace{1cm} We would like to thank Ruth Gregory and Harvey Reall
for correspondence and Chris Pope for finding an error in an earlier version.
SAH is funded by the Sims Scholarship and would like to
thank James Lucietti, Oisin MacConamhna and Toby Wiseman for discussions.

\appendix

\section{Scalar and vector modes}

We argue here that if the Einstein base manifold $B$ is compact, Riemannian and with
$\e=1$, then the scalar and vector modes will not produce instabilities in
the spacetimes under consideration. In particular, this covers the generalised
Schwarzschild-Tangherlini(-AdS) cases in which we found a criterion for
instability from the tensor modes.

The argument has two steps. Firstly, we will recall \cite{dfghm} that the scalar and vector
second order differential operators on $B$ are bounded below by their minimum
eigenvalue on the sphere $S^d$. Secondly, we will show that the
equations of motion for these modes depend on the base manifold only
through these differential operators. We will use $\tilde{h}$ and
$\tilde{h}_{\a}$ to denote generic scalar and vector modes on the base
manifold.

The second order differential operator on scalars is
$\tilde{\D}_S \tilde{h} = -\tnabla^{\a}\tnabla_{\a} \tilde{h}$. This is non-negative on compact manifolds
without boundary. There is always a zero mode corresponding to a
constant field. In particular, zero is the minimum eigenvalue on $S^d$
and also on any other such manifold $B$.

The second order differential operator on vectors in $B$ is
$(\tilde{\D}_V \tilde{h})_{\b} = -\tnabla^{\a}\tnabla_{\a}
\tilde{h}_{\b} + (d-1) \tilde{h}_{\b}$. By considering
\be
\int_B (\tnabla^{\a} \tilde{h}^{\b} + \tnabla^{\b} \tilde{h}^{\a})
(\tnabla_{\a} \tilde{h}_{\b} + \tnabla_{\b} \tilde{h}_{\a}) \geq 0 ,
\end{equation}
an integration by parts shows that for all eigenvalues $\lambda$ of $\tilde{\D}_V$, we have
$\lambda \geq 2(d-1)$. We have used the fact that the base manifold is Einstein with the same scalar
curvature as the unit sphere of appropriate dimension.
The inequality is saturated by modes that are Killing vectors.
It is well known that the minimum eigenvalue of $\tilde{\D}_V$ on $S^d$ is precisely $2(d-1)$
\cite{hig}. Therefore
\be
\lambda_{min}^B \geq \lambda_{min}^{S^d} .
\end{equation}

The equations for the various modes are found by expanding the
linearised equations
\be\label{eq:eomh}
\frac{1}{2} (\tilde{\D}_L h)_{ab} = c(d+1) h_{ab} ,
\end{equation}
into harmonics on the base manifold and considering each linearly
independent term separately. The various components of (\ref{eq:eomh})
may be classified by the number of free indices on the base manifold:
two, one or none. Each equation can then be written so that it is
tensorial on the base manifold. There are not many tensors
available that are linear in the perturbation, they are shown in table 4.

\begin{table}[h]
{\bf Table 4: } Available tensors on the base manifold. \\
  \begin{tabular}{|c|c|c|c|} \hline
Free base indices & tensors & vectors & scalars \\ \hline \hline
Two & $\tilde{h}_{\a\b}$ and $(\tilde{\D}_L \tilde{h})_{\a\b}$ &
  $\tnabla_{(\a}\tilde{h}_{\b)}$ &
  $\tnabla_{(\a}\tnabla_{\b)}\tilde{h}$ and $\tilde{g}_{\a\b}
  \tilde{h}$ \\ \hline 
One & - & $\tilde{h}_{\a}$ and $(\tilde{\D}_V \tilde{h})_{\a} $ &
  $\tilde{\partial}_{\a}\tilde{h}$  \\ \hline
None & - & - & $\tilde{h}$ and $\tilde{\D}_S \tilde{h}$ \\ \hline
  \end{tabular}
\end{table}

Multiplying these tensors in each equation will be differential
expressions involving functions of $r$ and $t$, but the $t$ dependence is
just an exponential. The equations
corresponding to the terms along the diagonal of table 4 will be of
the form
\be
F\left[\partial_r^2 \p^I,\partial_r \p^I,\p^I \right] \tilde{\D} \tilde{h} +
G\left[\partial_r^2 \p^I,\partial_r \p^I,\p^I \right] \tilde{h} = 0 ,
\end{equation}
where $F,G$ are functionals of the radial functions
$\p^I(r)$. Here $I$ indexes the scalar or vector modes because there
is more than one mode of each type in the decomposition of $h_{ab}$. In this
equation $\tilde{h}$ and $\tilde{\D}$ represent a harmonic mode on the
base and the corresponding differential operator. But we have
decomposed into harmonics of these differential operators, so
$\tilde{\D} \tilde{h} =
\lambda \tilde{h}$. This then gives a differential equation for the
$\p^I(r)$ that only depends on the base manifold through the
eigenvalue $\lambda$
\be
F\left[\partial_r^2 \p^I,\partial_r \p^I,\p^I \right] \lambda +
G\left[\partial_r^2 \p^I,\partial_r \p^I,\p^I \right]  = 0 .
\end{equation}

Above the diagonal in table 4 we see that there is only one possible
term in each case, after noting that $\tnabla_{(\a}\tnabla_{\b)}\tilde{h}$
does not arise from (\ref{eq:eomh}). Therefore the corresponding
equations will be of the form
\be
H\left[\partial_r^2 \p^I,\partial_r \p^I,\p^I \right] \tnabla
\tilde{h} = 0 ,
\end{equation}
where $\tnabla \tilde{h}$ is one of the tensors from above the
diagonal in table 4. But this simply implies the differential equation
$H=0$, with no dependence on the base manifold. Therefore the
equations for the perturbations only depend on the base manifold
through the eigenvalues of the second order differential operators on
the base.

As we saw for the tensor mode, instabilities are associated with the
eigenvalue being less than some critical value. This is the generic
situation and should be thought of as being due to a sufficiently
negative mass squared. However, in the first part of this appendix we saw that the
eigenvalues for the scalar and vector perturbations were bounded below
by the minimum eigenvalues on the sphere of the given dimension and
unit radius, $S^d$. It is expected, but to our knowledge not proven, that the standard
Schwarzschild-Tangherlini and Schwarzschild-AdS black holes are stable
against these vector and scalar perturbations
and therefore these eigenvalues must be
larger than any critical value. This suggests that the vector and
scalar modes do not contribute towards the instability of the
generalised Schwarzschild-Tangherlini and Schwarzschild-AdS black
holes that we have been considering in this work. It would be nice,
however, to check this explicitly from a perturbation analysis.

\section{Transverse tracefree gauge}

He we show that one may impose the tracefree condition in addition to the transverse
condition for vacuum spacetimes with a cosmological constant. This extends a familiar
argument to the case of nonzero cosmological constant.
Suppose the background spacetime in $d+2$ dimensions satisfies the {\it vacuum} field equations, possibly
with a cosmological constant,
\be
R_{ab} = c (d+1) g_{ab} ,
\end{equation}
then considering perturbations that are transverse in the sense of (\ref{eq:trans}) one may take the trace of
the perturbed equations
\be\label{eq:hpert}
\frac{1}{2} (\D_L h)_{ab} = c (d+1) h_{ab}\quad\Rightarrow\quad \square h^a{}_a + 2 c(d+1) h^a{}_a = 0 . 
\end{equation}
Consider solving the residual gauge freedom equations (\ref{eq:resid})
subject to the following initial conditions on
the spacelike hypersurface, $t=t_0$,
\bea\label{eq:init}
2 (\nabla^0 \xi_0 + \nabla^m \xi_m) & = & - h^a{}_a ,\nonumber \\
2 (-\nabla_m \nabla^m \xi_0 + \nabla_m \nabla_0 \xi^m-2c(d+1)\xi_0) & = & - \nabla_0 h^a{}_a ,
\eea
where $m$ runs over the spatial indices.
These equations may always be solved for $\xi$ and $\frac{d\xi}{dt}$ on the hypersurface.
This initial data then defines a vector field $\xi$ in the causal future of the hypersurface
using the residual gauge equation (\ref{eq:resid}). We can show that this vector field gives
a gauge transformation which sets the trace to zero. Define $f=h^a{}_a+2\nabla^a \xi_a$.
The initial conditions (\ref{eq:init}) and (\ref{eq:resid}) are seen to imply that $f=\frac{df}{dt}=0$ at
$t=t_0$. Now find the equation satisfied by $f$, using (\ref{eq:hpert}) and (\ref{eq:resid})
\bea
\square f & = & \square h + 2 \square \nabla^a \xi_a \nonumber \\
& = & - 2 c (d+1) h^a{}_a + 2 \nabla^a \square \xi_a - 2 c (d+1) \nabla^a \xi_a \nonumber \\
& = & - 2 c (d+1) \left[h^a{}_a + 2 \nabla^a \xi_a \right] = - 2 c (d+1) f .
\eea
We see that, perhaps unexpectedly, the equation for $f$ has no source term. This
fact implies that the vanishing
initial conditions force $f$ to vanish identically in the causal future of the initial hypersurface.
Issues of global hyperbolicity of the initial spacetime should not concern us as we are only
interested in the future development of the perturbation.
Thus we have shown that there exists a $\xi$ satisfying the residual gauge equation (\ref{eq:resid})
such that the trace of the perturbation may be set to $0$.

\section{No pure gauge solutions}

We show here that the tensor perturbations considered in this paper cannot be
pure gauge. Suppose there were a pure gauge mode, $h_{ab}=\nabla_a \xi_b + \nabla_b \xi_a$.
For the perturbations we are considering about the metric (\ref{eq:metric}), consider
the vanishing components
\be
h_{11} = 2(\partial_r \xi_1 - \frac{g^{\prime}(r)}{2 g(r)} \xi_1) = 0 .
\end{equation}
Therefore, noting that the form of the perturbation (\ref{eq:mode}) specifies the time dependence,
\be\label{eq:xi1}
\xi_1 = g(r)^{1/2} e^{\w t} \tilde{\xi}_1(\tilde{x}).
\end{equation}
Now consider the vanishing component
\be
h_{00} = 2 (\partial_t \xi_0 - \frac{f^{\prime}(r)}{2 g(r)} \xi_1 ) = 0 .
\end{equation}
Together with the form we found for $\xi_1$
this then implies that
\be
\xi_0 = \frac{f^{\prime}(r)}{2\w g(r)^{1/2}} e^{\w t}
\tilde{\xi}_1(\tilde{x}) .
\end{equation}
Next consider the vanishing component
\be
h_{01} = \partial_t \xi_1 + \partial_r \xi_0 - \frac{f^{\prime}(r)}{f(r)} \xi_0 = 0 .
\end{equation}
If we now substitute the expressions we have for $\xi_0, \xi_1$ into this equation,
we obtain a differential equation for $g(r)$ and $f(r)$ that is manifestly not satisfied by any
of the functions we used in this work. This implies that $\xi_0 = \xi_1 = 0$.
Now considering the vanishing components
\be
h_{0\a} = \partial_t \xi_{\a} + \partial_{\a} \xi_0 = 0 ,
\end{equation}
we see that $\xi_0=0$ implies that $\xi_{\a}$ is independent of $t$. But
if $\xi_{\a}$ is nonzero then it must depend on $t$ as $e^{\w t}$ in order for
the $h_{\a\b}$ term to have required time dependence. The conclusion is thus that
$\xi_a =0$ and therefore the perturbations we are considering can
never be pure gauge.

In the section on time dependent metrics we took $f=0$ and the metric
had no $t$ components. The argument of the previous paragraph will not
work in this case. The expression for $\xi_1$ (\ref{eq:xi1}) is now
just $\xi_1 = g(r)^{1/2} \tilde{\xi}_1(\tilde{x})$. We can
substitute this into the component
\be
h_{1\a} = \partial_r \xi_{\a} + \partial_{\a} \xi_1 -
\frac{2}{r}\xi_{\a} = 0 .
\end{equation}
Solving for $\xi_{\a}$ gives
\be\label{eq:defK}
\xi_{\a} = \left[ - r^2\int^r \frac{g(s)^{1/2}}{s^2} ds
\right] \partial_{\a} \tilde{\xi}_1(\tilde{x}) \equiv
K(r)\partial_{\a} \tilde{\xi}_1(\tilde{x}) .
\end{equation}
Finally, substitute these results into the remaining, nonvanishing, component
and recall the form of the perturbation (\ref{eq:tpert}) and the
definition of $K(r)$ in (\ref{eq:defK})
\bea
h_{\a\b} & = & \tilde{\nabla}_{\a} \xi_{\b} + \tilde{\nabla}_{\b}
\xi_{\a} + \frac{2r}{g(r)} \tilde{g}_{\a\b}(\tilde{x}) \xi_1 \nonumber \\
 & = & 2 K(r) \tilde{\nabla}_{\a} \partial_{\b} \tilde{\xi}_1(\tilde{x}) +
\frac{2r}{g(r)^{1/2}} \tilde{g}_{\a\b}(\tilde{x}) \tilde{\xi}_1(\tilde{x}) \nonumber \\
 & = & r^2 \p(r) \tilde{h}_{\a\b}(\tilde{x}) .
\eea
For the last two lines to be equal, we must have either $K(r)
\propto \frac{r}{g(r)^{1/2}}$ or
$h_{\a\b} \propto \tilde{g}_{\a\b} \tilde{\xi}_1
\propto \tilde{\nabla}_{\a} \partial_{\b} \tilde{\xi}_1$.
The former possibility is not true for the functions $g(r)$ that we have
been considering whilst the latter is not consistent with
$h^{\a}{}_{\a} = 0$. Therefore there are no pure gauge solutions. We
could have used this argument in the previous case also, but the
argument we gave instead did not use the tracefree property at any point.

\end{document}